\documentclass[aps,prl,superscriptaddress,twocolumn]{revtex4}
\usepackage{graphicx,amssymb,amsmath,color}
\usepackage{xcolor}
\usepackage[pdfstartview=FitH,colorlinks=true,citecolor=blue,linkcolor=blue,urlcolor=blue]{hyperref}

\usepackage{nccmath}
\usepackage{comment}

\begin{document}
\title{Interaction induced Anderson transition in a kicked one dimensional Bose gas}
\author{H. Olsen}
\affiliation{Universit\'e C\^ote d'Azur, CNRS, Institut de Physique de Nice, France}
\author{P. Devillard}
\affiliation{Aix Marseille Univ, Universit\'e de Toulon, CNRS, CPT, Marseille, France}
\author{G. Aupetit-Diallo}
\affiliation{SISSA, Via Bonomea 265, I-34136 Trieste, Italy}
\author{P. Vignolo}
\affiliation{Universit\'e C\^ote d'Azur, CNRS, Institut de Physique de Nice, France}
\affiliation{Institut Universitaire de France}
\author{M. Albert}
\affiliation{Universit\'e C\^ote d'Azur, CNRS, Institut de Physique de Nice, France}
\affiliation{Institut Universitaire de France}

\begin{abstract}
  We investigate the Lieb-Liniger model of one-dimensional bosons subjected to periodic kicks. In both the non-interacting and strongly interacting limits, the system undergoes dynamical localization, leading to energy saturation at long times. However, for finite interactions, we reveal an interaction-driven transition from an insulating to a metallic phase at a critical kicking strength, provided the number of particles is three or more. Using the Bethe Ansatz solution of the Lieb-Liniger gas, we establish a formal correspondence between its dynamical evolution and an Anderson model in $N$ spatial dimensions, where $N$ is the number of particles. This theoretical prediction is supported by extensive numerical simulations for three particles, complemented by finite-time scaling analysis, demonstrating that this transition belongs to the orthogonal Anderson universality class.
\end{abstract} 

\maketitle


The phenomenon of Anderson localization, originally introduced in the context of electron transport in disordered media \cite{Anderson}, is now recognized as a universal feature of classical and quantum wave dynamics. In non-interacting systems, interference effects lead to the exponential localization of eigenstates and the suppression of diffusion. A key aspect of Anderson localization is the existence of a disorder-induced metal-insulator transition in three dimensions (3D), where a critical level of disorder separates a diffusive metallic phase at small disorder from a localized insulating phase at strong disorder \cite{Thouless}. This transition, known as the Anderson transition, shares key properties with second-order phase transitions and can be described within the framework of the one-parameter scaling theory \cite{gangof4}. According to this theory, the localization length diverges at criticality as $\ell\sim (W-W_c)^{-\nu}$, where $W$ is the disorder strength, $W_c$ the critical disorder, and $\nu$ the critical exponent. On the other side of the transition this is the diffusion constant $D$ which vanishes as $(W_c-W)^{s}$. Importantly, Anderson localization is fundamentally dependent on dimensionality: while all states are localized in one dimension (1D) for any amount of disorder, in 3D (and for larger dimensions), a mobility edge separates localized and extended states \cite{gangof4}. A major open question is how localization is affected by interactions, particularly in low-dimensional and driven systems. While interactions generally promote delocalization \cite{GiamarchiSchulz}, strong disorder can stabilize a many-body localized (MBL) phase, which exhibits ergodicity breaking and prevents thermalization \cite{BaskoAleinerAltshuler,NandkishoreHusereview,AletLaflorenciereview,Ponte2015,Ponte2015b}.

Dynamical localization, the analog of Anderson localization for driven systems, occurs in periodically kicked quantum systems, where momentum states play the role of spatial sites in the Anderson model \cite{FishmanGrempelPrange,GrempelPrangeFishman,Shepelyansky1986}. The quantum kicked rotor (QKR) is a paradigmatic model in this context: in the absence of interactions, dynamical localization manifests itself as an energy saturation at long times due to destructive quantum interference. The QKR has been extensively studied in atomic experiments, leading to key insights into Anderson localization \cite{MooreRobinsonBharuchaetal1995}, including the observation of Anderson transitions \cite{Chabeetal2008,Lemarie2009,WangGarcia2009,LemarieLignieretal2010,Lopez2012} and the effects of symmetries on weak localization \cite{Hainaut2018,Lemarie2017}.

Ultracold atomic gases provide an ideal platform \cite{Lewenstein2007,BlochDalibardZwerger,Lewenstein2012} to investigate the interplay between interactions and localization in driven quantum systems. One-dimensional (1D) Bose gases, described by the Lieb-Liniger model \cite{LiebLiniger}, are particularly well suited for exploring these effects, as they allow for precise control over interactions and can be routinely realized in experiments \cite{Paredesetal,Kinoshita2004,Kinoshita2005,CazalillaCitroGiamarchiOrigancRigol}. Despite recent progress, the fate of dynamical localization in the presence of interactions remains a subject of debate. In the non-interacting and strongly interacting (Tonks-Girardeau) limits \cite{Tonks,Girardeau}, dynamical localization persists for any number of particles \cite{RylandsRozenbaumGalitskiKonik,Vuatelet2021}. For two interacting bosons, localization is also always preserved \cite{ChicireanuRancon}. However, for finite interactions and more than two particles, theoretical and experimental results are conflicting. Mean-field studies suggest that interactions induce subdiffusive energy growth, destroying dynamical localization \cite{Shepelyansky1993,Pikovsky2008,Flach2009,Flach2011,Cherroret2014,Lellouch2020}, whereas, Rylands et al. argued that dynamical localization persists for an arbitrary number of particles in the weak-kicking regime \cite{RylandsRozenbaumGalitskiKonik}. Experimental studies have also provided contradictory conclusions: two experiments found that interactions destroy localization \cite{CaoetalSantaBarbara,Gupta2021}, while another observed its robustness \cite{Nagerl2023}.

In this letter, we resolve this open problem by employing exact methods. We study the kicked Lieb-Liniger model, describing a 1D Bose gas subjected to periodic kicks, and establish a formal correspondence between its dynamical evolution and an Anderson model in $N$ spatial dimensions, where $N$ is the number of particles. This mapping enables us to predict the existence of an Anderson-type metal-insulator transition at a critical kicking strength, dependent on interactions for $N\ge 3$. Our findings are supported by extensive numerical simulations and finite-time scaling analysis for $N=3$, providing new insights into the fundamental interplay between interactions and localization in driven quantum systems.


\textit{Model.}-- 
We consider $N$ identical bosons of mass $m$ and coordinates $y_i$, living on a ring of size $L$ with point-like repulsive interactions of strength $g$, and periodic boundary conditions. In addition, the atoms are subjected to a kick potential of the form  $\mathcal{K}\cos(2\pi y_i/L)\sum_n \delta (t'-n\tau)$ with period $\tau$ and amplitude $\mathcal{K}$. 
After rescaling time with $\tau$, positions with $\ell=L/2\pi$ and energy with $\varepsilon=m\ell^2/\tau^2$ the two contributions of the Hamiltonian at which the atoms are subjected read

\begin{equation}\label{eq_H_LL}
  H_0 = \sum_{i=1}^N \frac{p_i^2}{2}  +  c \, \sum_{i>j} \delta(x_i-x_j),
\end{equation}
\begin{equation}\label{eq_H_K}
  H_{\rm{kick}}=H_K \sum_{n=-\infty}^{+\infty}\delta(t-n)\,,\quad H_K=K \sum_{i=1}^N\cos(x_i),
\end{equation}
where $c=g/(\varepsilon\ell)$, $K=\mathcal{K}/(\tau\varepsilon)$, $t=t'/\tau$, $x_i=y_i/\ell$, $p_i=-i\hbar_e \partial / \partial x_i$ the momenta of the particles satisfying the commutation relation $[x_j,p_j]=i \hbar_e$, where 
$\hbar_e=\hbar/(\varepsilon\tau)$ is the effective Planck's constant. $H_0$ is the Lieb-Liniger Hamiltonian which can be exactly solved with Bethe Ansatz \cite{LiebLiniger}. The eigenstates are labeled by a set of $N$ rapidities $\lambda_i$ and can be written, in the fundamental sector $x_1\le x_2\le ...\le x_N$, as

\begin{equation}\label{eq_psi}
  \Psi_{\{\lambda_j\}}(\{x_j\}) = \sum_{P\in S_N} A_P  \exp\Bigl(i \sum_{k=1}^N \lambda_{P(k)} x_k\Bigr),
\end{equation}
where $S_N$ is the permutation group of $N$ elements. The $A_P$ are coefficients given by
\begin{equation}\label{eq_ap}
  A_P = \mathcal N (-1)^{{\rm sgn}(P)}  \prod_{1\le k < j \le N}[\lambda_{P(j)} - \lambda_{P(k)}-ic]
\end{equation}
with $\mathcal N$ the normalization constant. In addition, the rapidities are all real and fulfill the $N$ coupled Bethe equations
\begin{equation}\label{eq_bethe}
  \lambda_j = I_j - \frac{1}{\pi} \sum_{k=1}^N \textrm{arctan}\left(\frac{\lambda_j - \lambda_k}{c}\right),
\end{equation}
where the Bethe numbers $I_j$ are integers for odd $N$ and half-integers for even $N$, and must all be distinct. The energy of a Bethe state is given by $E_{\vec \lambda} = \frac{\hbar_e^2}{2} \sum_{i=1}^N \lambda_i^2$ and its momentum $P_{\vec \lambda} = \hbar_e \sum_{i=1}^N \lambda_i$. The ground state corresponds to the lowest available set of Bethe numbers. For instance, for $N=3$, it is given by $\vec I=(-1,0,+1)$.
Due to the bosonic statistics, these states are symmetric under permutations of spatial coordinates and rapidities, allowing us to restrict our calculations to the fundamental rapidity sector $I_1<I_2<...<I_N$. It is important to note that the rapidities $\lambda_i$ differ from the actual momenta of the particles except in the non-interacting limit. In the limit of infinite interactions ($c \to +\infty$), the rapidities coincide with the momenta of the corresponding free fermions via the Bose-Fermi mapping \cite{Girardeau}. In general, however, they do not reside on a regular lattice (which would have spacing $ 2\pi \ell/L = 1$ in our units), but are instead coupled through the Bethe equations \eqref{eq_bethe}.  

Before proceeding further, we summarize a few key results relevant to our study. In the non-interacting limit ($c=0$), the system reduces to $N$ independent quantum kicked rotors, each of which undergoes dynamical localization, characterized by energy saturation at long times \cite{GrempelPrangeFishman}. At short times, the momentum-space dynamics of each rotor is diffusive, in agreement with the classical description, with a diffusion constant approximately given by $D\simeq K^2/4$. However, beyond the localization time $\tau_{loc}\simeq K^2/4\hbar_e^2$, quantum interference effects halt diffusion, leading to energy saturation at a characteristic momentum scale $p_{loc}\simeq K^2/4\hbar_e$, the localisation momentum \cite{Shepelyansky1986}. In the opposite limit of infinite interactions ($c \to +\infty$), the system maps onto a gas of free fermions via the Bose-Fermi correspondence \cite{Girardeau}. In this regime, the localization properties remain qualitatively similar to the non-interacting case, as previously demonstrated in theoretical and numerical studies \cite{RylandsRozenbaumGalitskiKonik,Vuatelet2021}. Additionally, the case of two interacting bosons ($N=2$) has been investigated for arbitrary interaction strength \cite{ChicireanuRancon}, revealing that dynamical localization persists regardless of the interaction strength.


\textit{General mapping to an Anderson-like model.}--
Since the Hamiltonian is time-periodic, the system's evolution can be described by iterating the time evolution over one period (here and in the following $\tau=1$)

\begin{equation}
  \label{eq_Ufloquet}
  U=e^{-i H_0/\hbar_e} e^{-i H_K/\hbar_e}.
\end{equation}
For $N=1$ it is well known that the eigenstates of this Floquet operator are dynamically localized in momentum space \cite{GrempelPrangeFishman}. This localization can be understood through a formal analogy between the Floquet eigenvalue equation and a one-dimensional Anderson model. A natural question then arises: to what extent do these results generalize  to a larger number $N$ of interacting particles? Here, we demonstrate that a similar formal mapping can be established in the $N$-dimensional rapidity space. Instead of relying on the tangent map approach \cite{FishmanGrempelPrange,GrempelPrangeFishman}, we follow the strategy outlined in Ref. \cite{RossiniFaziogroup}. Our basis of states consists of the Bethe states $\vert {\vec \lambda} \rangle$, rather than the usual momentum eigenstates. The Floquet eigenstates of the evolution operator \eqref{eq_Ufloquet} satisfy $U|\phi_\alpha\rangle=e^{-i \frac{\mu_\alpha}{\hbar_e}} |\phi_\alpha\rangle$, where $\mu_{\alpha}$ are the Floquet quasi-energies. Moving to the interaction representation by defining $|{\tilde  \phi}_{\alpha}\rangle = e^{iH_0/ 2\hbar_e} |\phi_{\alpha}\rangle$, we note that this unitary transformation does not alter the state structure in rapidity space, since $H_0$ is diagonal in this basis. The Floquet equation then takes the form
\begin{equation}
\sum_{{\vec \mu} \neq {\vec \lambda}} \! W_{{\vec \lambda},{\vec \mu}}  \langle {\vec \mu} \vert {\tilde \phi}_{\alpha}\rangle    +   W_{{\vec \lambda},{\vec \lambda}}  
\langle {\vec \lambda} \vert {\tilde \phi}_{\alpha}\rangle  =
 2 \cos\Bigl(\frac{\mu_{\alpha}}{\hbar_e}\Bigr)  \langle {\vec \lambda} \vert {\tilde \phi}_{\alpha}\rangle,
\label{FloquetEq9}
\end{equation}
with the matrix elements
\begin{eqnarray}
 W_{{\vec \lambda},{\vec \mu}}  &=& 2 \, {\rm Re} \biggl\lbrace \exp\Bigl(-i \frac{ E_{\vec \lambda}+E_{\vec \mu}}{2\hbar_e}  \Bigr) M_{{\vec \lambda},{\vec \mu}}\biggr\rbrace, \\
M_{{\vec \lambda},{\vec \mu}}  &=& \langle{\vec \lambda} \vert e^{-iH_K/\hbar_e} \vert {\vec \mu} \rangle,
\end{eqnarray}
and ${\rm Re}\lbrace X \rbrace$ denotes the real part of a complex number $X$. This equation closely resembles an Anderson model with long-range hopping in an $N$-dimensional rapidity space. However, for this analogy to hold rigorously, two key conditions must be met. First, the on-site energy $W_{\vec \lambda, \vec \lambda}$ must exhibit pseudo-random behavior, which is ensured if $\hbar_e$ is incommensurate with $4\pi$ \cite{FishmanGrempelPrange}. Second, the hopping terms must decay faster than $1/||\vec\lambda-\vec\mu||$ \cite{Mirlin1996}. This requirement is satisfied in our model since the matrix elements $M_{{\vec \lambda},{\vec \mu}}$ decay exponentially \cite{suppmat}.

This formal correspondence leads to a highly nontrivial prediction: an Anderson-type insulator-to-metal transition should emerge for $N\ge 3$. This, in particular, explains why no transition was observed in the two-particle case studied in Ref. \cite{ChicireanuRancon}. However, several subtleties need to be addressed. Unlike a standard Anderson model describing a particle on a $d$-dimensional lattice, our system prohibits double occupancy of a site, as two rapidities cannot be equal. Additionally, the many-body wave function is invariant under rapidity permutations, allowing us to restrict the problem to the fundamental sector $\lambda_1<\lambda_2<...<\lambda_N$. Moreover, disorder in our model is genuinely $N$-dimensional only when interactions are finite. In both the non-interacting limit ($c\to 0$) and the Tonks-Girardeau limit ($c\to \infty$), the Bethe equations \eqref{eq_bethe} decouple, meaning that all rapidities evolve independently \cite{suppmat}. In these limits, the system effectively reduces to $N$ independent copies of the one-dimensional kicked rotor, which explains why dynamical localization persists. By contrast, for finite $c$, the rapidities are coupled, and the disorder acquires an $N$-dimensional structure due to the non-factorization of the evolution operator. 

Under these conditions, we expect an Anderson transition at a critical stochasticity parameter $K_c$, which depends on both the interaction strength $c$ and the effective Planck constant $\hbar_e$. Furthermore, we anticipate that $K_c$ diverges in both the weakly interacting ($c\to 0$) and strongly interacting ($c\to \infty$) limits, reinforcing the absence of a transition in these regimes.


\textit{Results for $N=3$ bosons.}--
To illustrate our findings, we consider the case of $N=3$ bosons, where extensive numerical calculations can be performed. To do so, we truncate the Hilbert space by restricting the Bethe numbers to the range $-N_s\le I_j\le N_s$, where $N_s$ is an integer (since $N=3$ is odd). The computation of the Floquet operator matrix elements \eqref{eq_Ufloquet} then reduces to evaluating $M_{{\vec \lambda},{\vec \mu}}  = \langle{\vec \lambda} \vert e^{-iH_K/\hbar_e} \vert {\vec \mu} \rangle$ since $e^{-i H_0/\hbar_e}$ is diagonal in this basis. These matrix elements can be computed analytically using \eqref{eq_psi} \cite{suppmat}.
The total matrix size in the truncated basis is given by $N_M=\binom{2N_s+1}{3}$. We initialize the system in the ground state, characterized by Bethe numbers $\vec I=(-1,0,1)$, and evolve it numerically by iterating the Floquet operator in the truncated Hilbert space with $N_s=29$ for selected values of $K$, $c$, and $\hbar_e$. The total energy is then computed as $\langle E(t)\rangle = \sum_{\vec \lambda} |\alpha_{\vec \lambda}(t)|^2 E_{\vec\lambda}$, where $\alpha_{\vec \lambda}(t)$ represents the amplitude of the many-body wave function in the Bethe state $|\vec\lambda\rangle$.

\begin{figure}
  \includegraphics[width=0.88\linewidth]{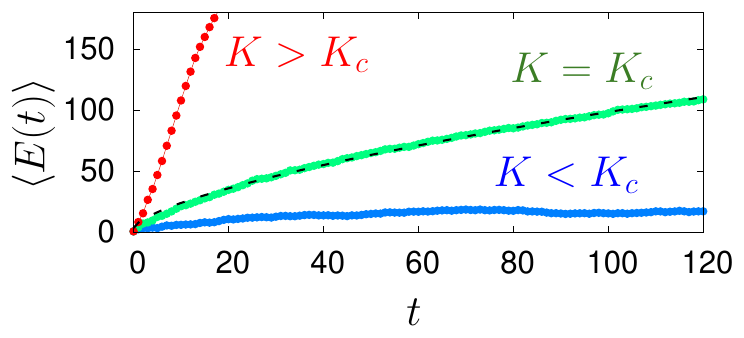}\\
  \includegraphics[width=0.88\linewidth]{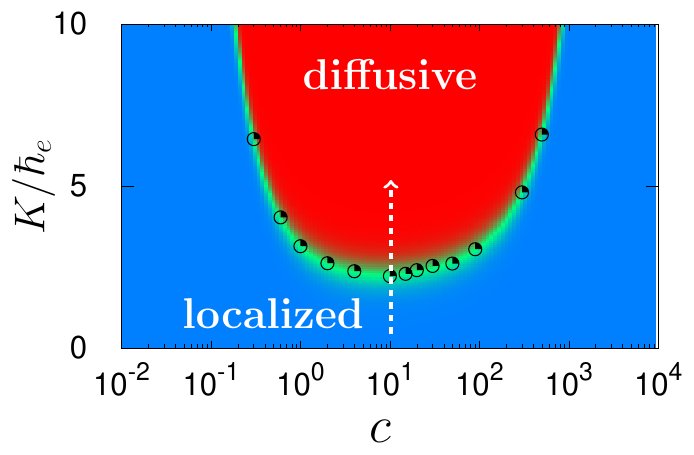}
  \caption{\label{fig_energy} Upper panel: time evolution of the total energy of the three particles for different values of the stochasticity parameter $K$ at finite interaction strength $c=10$. Close to the critical point $K_c/\hbar_e\simeq 2.1$ (green curve) anomalous diffusion is observed with exponent $2/3$ (black dashed line). For $K=1.5\hbar_e<K_c$ (blue curve) the energy growth saturates as expected in the localized regime. Above the critical point $K=3.2\hbar_e>K_c$ (red curve) the dynamics is diffusive up to some time where the evolution is sensitive to finite size effects. Lower panel: phase diagram of the dynamical phases in the $c$-$K$ plane. The circles correspond to numerical determination of the critical line while the color code is just a guide to the eyes based on a fit of the data. Here $\hbar_e=2.89$, $N=3$ and $N_s=29$. The white arrow indicates the path we follow to analyse the transition at $c=10$.}
\end{figure}

This approach successfully reproduces the expected behavior in both the non-interacting and strongly interacting limits. It also enables us to compute the phase diagram shown in Fig. \ref{fig_energy}, which demonstrates the existence of a metal-insulator phase transition at a critical value of $K$ that depends on the interaction strength and diverges in both previously mentioned limits. In the following, we explain how this phase diagram was constructed and focus on a specific interaction strength, $c=10$, where the critical value of $K$ remains small enough to allow for accurate numerical calculations despite our limited system size.

As a first result, we present the energy dynamics in Fig. \ref{fig_energy} for three different values of the stochasticity parameter $K$. This provides an initial indication of the phase transition, as we identify three distinct behaviors. For $K<K_c$, the energy saturates at long times, a hallmark of dynamical localization. Conversely, for $K>K_c$, the energy grows unboundedly, exhibiting a linear increase, albeit with corrections due to finite-size effects. Between these two regimes, we identify a critical behavior at $K=K_c$, characterized by anomalous diffusion, with the energy scaling as $\langle E(t) \rangle\sim t^{2/3}$. This scaling is consistent with the predictions of the one-parameter scaling theory in three-dimensional Anderson transitions \cite{gangof4,Lemarie2009}. 

\begin{figure}
  \includegraphics[width=0.48\linewidth]{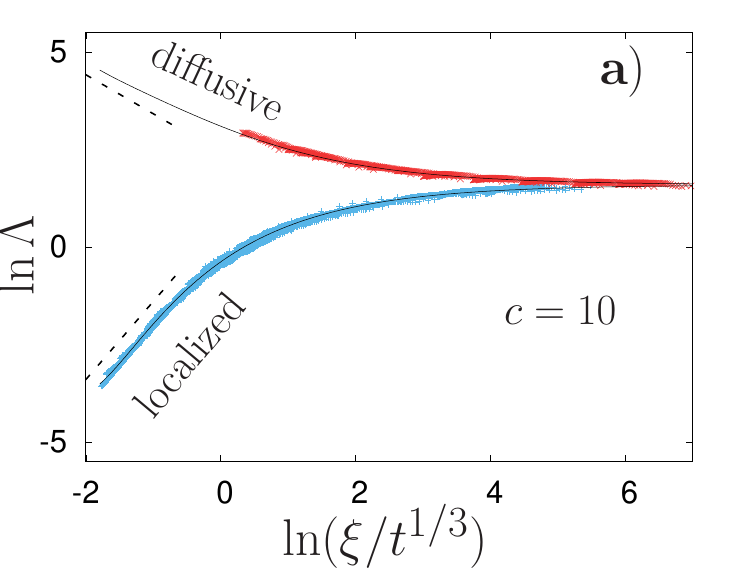}
  \includegraphics[width=0.48\linewidth]{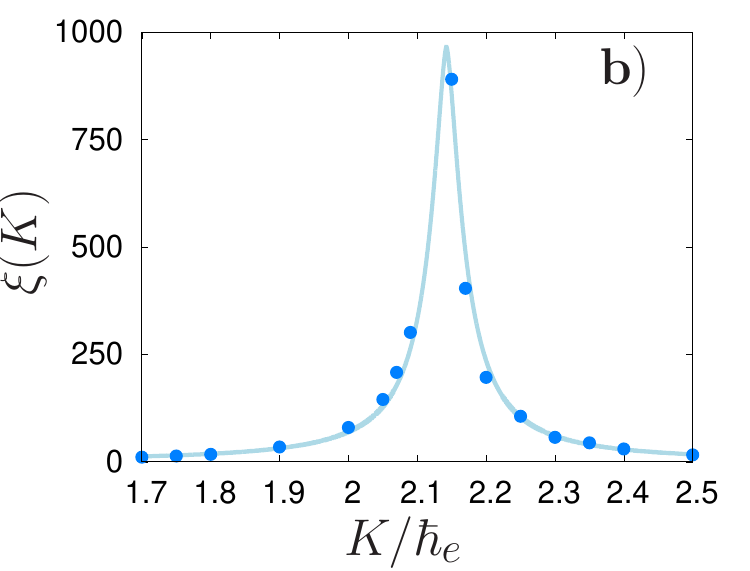}
  \caption{\label{fig_scaling}  Finite time scaling applied to the numerical results with $c=10$ and $N=3$. The time evolution of $\langle E \rangle$ is computed as a function of time, from 20 to 200 kicks, for several values of $K/\hbar_e$ between $K/\hbar_e=0.4$ to $K/\hbar_e=2.9$. The scaling function $\ln\Lambda$ (panel a), with $\Lambda=\langle E\rangle/t^{2/3}$, displays a lower branch (blue) associated with the localized regime and an upper branch (red) associated with the diffusive regime. The continuous curve is a fit using a Taylor expansion of the scaling function up to fourth order and a critical exponent $\nu=1.56$. The dependence of the scaling parameter $\xi(K)$ is displayed on panel b) and shows a divergent behavior around the critical point $K_c/\hbar_e=2.13$. The continuous light-blue curve is a fit using a critical exponent $\nu=1.56$ (see text).}
\end{figure}

We now proceed to a more precise characterization of the transition. Identifying a phase transition is challenging due to finite-time and finite-size effects. This difficulty arises because, at the transition, the localization momentum \( p_{\text{loc}} \) diverges and can significantly exceed the system size, which then acts as an upper bound for the effectively observed localization momentum. To circumvent this issue, Lemari\'e et al. \cite{Lemarie2009} proposed a finite-time scaling analysis inspired by the finite-size scaling method developed for the Anderson model \cite{Kramer1981,Pichard1981}. We closely follow this approach to demonstrate that the observed transition indeed belongs to the orthogonal Anderson universality class.  

To apply this method, we first recall some important theoretical results. In the localized regime (\( K < K_c \)), the energy saturates at long times as $\langle E\rangle \sim p_{\text{loc}}^2$. If an Anderson transition exists, we expect \( p_{\text{loc}} \) to diverge near the critical point as $p_{\text{loc}} \sim (K_c - K)^{-\nu}$. On the other side of the transition (\( K > K_c \)), the dynamics is diffusive, with the energy growing as $\langle E \rangle \sim D(K) t$ where the diffusion constant \( D(K) \) vanishes as \( (K - K_c)^s \). In three dimensions, it is known that \( s = \nu \) \cite{Wegner1976}. At the critical point (\( K = K_c \)), anomalous diffusion is expected, characterized by an exponent \( 2/3 \) \cite{Lemarie2009}. The one-parameter scaling theory postulates that the quantity  
\begin{equation}
  \Lambda(K,t) = \langle E(t)\rangle t^{-2/3} = f\left( \xi(K) t^{-1/3} \right),
\end{equation}  
where \( f \) is an arbitrary function, and the scaling parameter \( \xi(K) \) depends only on \( K \). To test this hypothesis, we present in Fig.~\ref{fig_scaling}a our numerical results obtained for \( c = 10 \), \( \hbar_e = 2.89 \), and several values of \( K \). Each dataset is shifted along the \( x \)-axis by a quantity \( \ln \xi(K) \) to align all the data points onto the same universal scaling curve. The values of \( \xi(K) \) are determined by minimizing the distance between the corresponding values of \( \ln(\xi(K)t^{-1/3}) \) for each value of \( \ln\Lambda \). Remarkably, all data collapse onto a single universal function, with a lower branch corresponding to the localized regime and an upper branch corresponding to the diffusive regime. Figure~\ref{fig_scaling}b displays the scaling parameter \( \xi(K) \), which diverges at \( K_c \) following a power law consistent with the prediction of the orthogonal Anderson universality class, namely \( \nu = 1.56 \). To model this divergence and account for finite-size effects, we fit our data using the function ${\xi(K)}^{-1} = \alpha |K - K_c|^\nu + \beta$.

\begin{figure}
  \includegraphics[width=0.59\linewidth]{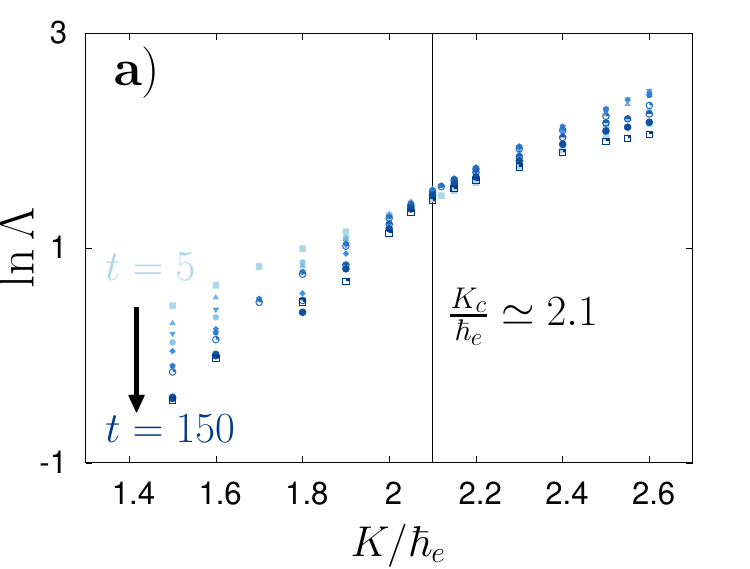}
  \includegraphics[width=0.37\linewidth]{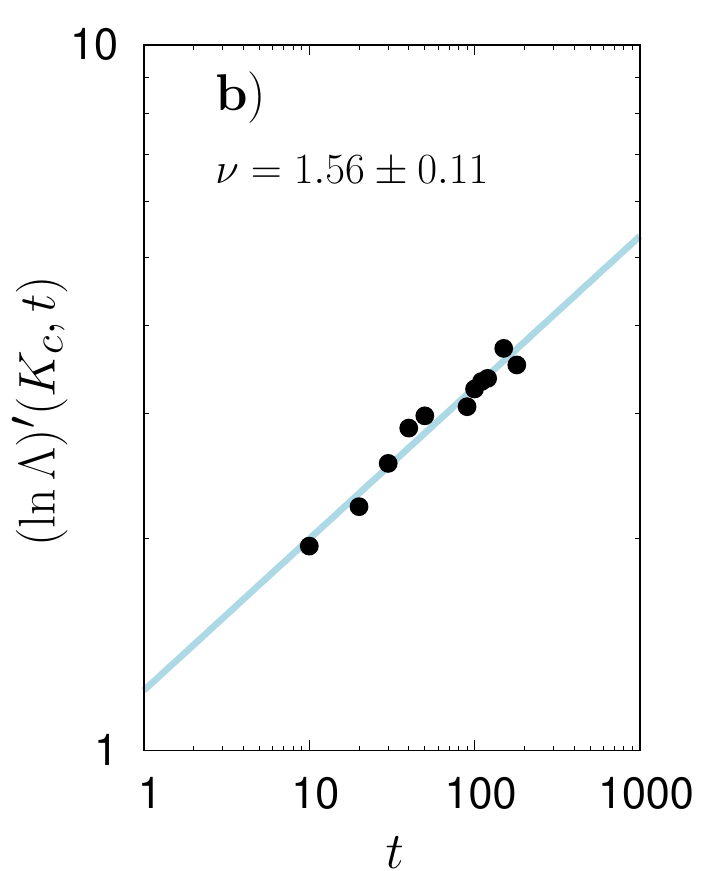}
  \caption{\label{fig_scaling2} (a) The rescaled quantity $\ln\Lambda=\ln[\langle E(t) \rangle t^{-2/3}]$ as a function of $K$ for different times between $t=5$ (light blue) and $t=150$ (dark blue). All curves intersect approximately at the critical point $(K_c/\hbar_e\simeq 2.1,\ln\Lambda_c\simeq 1.54)$ demonstrating the existence of a metal-insulator transition. (b) Determination of the critical exponent $\nu$ by fitting $(\ln\Lambda)'(K_c,t)\sim t^{1/3\nu}$ in log-log scale. Parameters are the same as in Fig. \ref{fig_scaling}.}
\end{figure}

Finally, to precisely locate the critical point and extract the critical exponent \( \nu \), we plot in Fig.~\ref{fig_scaling2}a \( \ln \Lambda(K,t) \) as a function of \( K \) for different time values. This method is particularly convenient since, at the critical point, \( \Lambda(K_c,t) \) is time-independent. We thus determine \( K_c \) by identifying the crossing points of all curves. Furthermore, we extract the critical exponent by analyzing the behavior of \( \ln \Lambda(K,t) \) near \( K_c \). Assuming \( \xi(K) \sim |K - K_c|^{-\nu} \), the scaling function depends on \( (K - K_c)t^{1/3\nu} \), leading to \( \ln \Lambda(K,t) \simeq \ln \Lambda_c + \mathcal{C}_1 (K - K_c)t^{1/3\nu} + \dots \). Thus, the slope of \( \ln \Lambda \) near \( K_c \) is proportional to \( t^{1/3\nu} \). By plotting this slope as a function of time on a log-log scale, we determine the critical exponent via a linear fit. As shown in Fig.~\ref{fig_scaling2}b, this yields \( \nu = 1.56 \pm 0.11 \), in agreement with the state-of-the-art value for the orthogonal Anderson transition \cite{SlevinOhtsuki1997}, \( \nu = 1.57 \pm 0.02 \). For comparison, the exponent for the unitary Anderson transition is \( \nu = 1.43 \pm 0.04 \) \cite{SlevinOhtsuki1997}.


\textit{Conclusion.}--
We have investigated the dynamics of a periodically kicked one-dimensional Bose gas with repulsive contact interactions. By establishing a formal correspondence between the system's evolution and an Anderson model in an $N$-dimensional rapidity space, we have demonstrated the existence of an interaction-driven metal-insulator transition for $N \geq 3$. Our extensive numerical simulations for three bosons confirm this transition and allow us to extract the critical stochastic parameter $K_c$ as a function of interactions as well as the associated critical exponent. 

This work, grounded in exact calculations, opens new perspectives on the interplay between interactions and Anderson localization. A key open question remains the reconciliation of our findings with conflicting experimental results \cite{CaoetalSantaBarbara,Gupta2021,Nagerl2023}. Additionally, the connection with mean-field approaches, which predict delocalization as soon as interactions are introduced, warrants further investigation. A deeper analysis of our effective model in the infinite-dimensional limit could provide crucial insights into this discrepancy.


\begin{acknowledgments}
  We would like to acknowledge helpful discussions with L. Bellando De Castro, J. C. Garreau, B. Gr\'emaud, F. Hébert, M. Landini, C. Miniatura, A. Ran\c con and F. Werner. This work has benefited from the financial support of Agence Nationale de la Recherche under Grant No ANR-21-CE47-0009 Quantum-SOPHA and Grant No ANR-23-PETQ-0001 Dyn1D France 2030.

  Note.—During the publication process, we found that A. Yang {\it et al} \cite{Landini2025} independently posted a preprint on the same topic at the same time.

\end{acknowledgments}


\bibliographystyle{apsrev4-2}

\end{document}


\title{Supplemental material: Interaction induced Anderson transition in a kicked one dimensional Bose gas}
\author{H. Olsen}
\affiliation{Universit\'e C\^ote d'Azur, CNRS, Institut de Physique de Nice, France}
\author{P. Devillard}
\affiliation{Aix Marseille Univ, Universit\'e de Toulon, CNRS, CPT, Marseille, France}
\author{G. Aupetit-Diallo}
\affiliation{SISSA, Via Bonomea 265, I-34136 Trieste, Italy}
\author{P. Vignolo}
\affiliation{Universit\'e C\^ote d'Azur, CNRS, Institut de Physique de Nice, France}
\affiliation{Institut Universitaire de France}
\author{M. Albert}
\affiliation{Universit\'e C\^ote d'Azur, CNRS, Institut de Physique de Nice, France}
\affiliation{Institut Universitaire de France}

\maketitle

\section{Bethe ansatz solution for three particles}
In this section, we recall the expressions of the eigenstates of Hamiltonian (1) from the main text for the case of $N=3$ identical bosons \cite{LiebLiniger,Lamacraft}. Following the Bethe Ansatz, the wavefunction in each position sector is expressed as a superposition of plane waves with fixed energy. In the fundamental position sector $x_1<x_2<x_3$, the Bethe wavefunction takes the form 

\begin{equation}
\begin{split}
    \Psi(x_1, x_2, x_3) &=  A_{123} e^{i(\lambda_1 x_1 + \lambda_2 x_2 + \lambda_3 x_3)} + A_{132} e^{i(\lambda_1 x_1 + \lambda_3 x_2 + \lambda_2 x_3)}  \\
           & + A_{213} e^{i(\lambda_2 x_1 + \lambda_1 x_2 + \lambda_3 x_3)} + A_{231} e^{i(\lambda_2 x_1 + \lambda_3 x_2 + \lambda_1 x_3)} \\
           & + A_{312} e^{i(\lambda_3 x_1 + \lambda_1 x_2 + \lambda_2 x_3)} + A_{321} e^{i(\lambda_3 x_1 + \lambda_2 x_2 + \lambda_1 x_3)},\\
           \end{split}
\end{equation}
where the positions $x_j$ are expressed in units of $\ell=L/(2\pi).$ 
The non-normalized amplitudes $A_{ijk}$, which depend on the rapidities $\lambda_j$ and the interaction strength $c$, are determined by enforcing the cusp condition at $x_i=x_j$. This accounts for the discontinuity in the derivative due to the contact interactions \cite{LiebLiniger,Lamacraft}. The amplitudes are given by

\begin{align*}
A_{123}(\lambda_1, \lambda_2, \lambda_3, c) &= (\lambda_1 - \lambda_2 + ic)(\lambda_1 - \lambda_3 + ic)(\lambda_2 - \lambda_3 + ic) \\
A_{132}(\lambda_1, \lambda_2, \lambda_3, c) &= (\lambda_1 - \lambda_2 + ic)(\lambda_1 - \lambda_3 + ic)(\lambda_2 - \lambda_3 - ic) \\
A_{213}(\lambda_1, \lambda_2, \lambda_3, c) &= (\lambda_1 - \lambda_2 - ic)(\lambda_1 - \lambda_3 + ic)(\lambda_2 - \lambda_3 + ic) \\
A_{312}(\lambda_1, \lambda_2, \lambda_3, c) &= (\lambda_1 - \lambda_2 + ic)(\lambda_1 - \lambda_3 - ic)(\lambda_2 - \lambda_3 - ic) \\
A_{321}(\lambda_1, \lambda_2, \lambda_3, c) &= (\lambda_1 - \lambda_2 - ic)(\lambda_1 - \lambda_3 - ic)(\lambda_2 - \lambda_3 - ic) \\
A_{231}(\lambda_1, \lambda_2, \lambda_3, c) &= (\lambda_1 - \lambda_2 - ic)(\lambda_1 - \lambda_3 - ic)(\lambda_2 - \lambda_3 + ic).
\end{align*}
Finally the rapidities $\lambda_j$ are obtained by imposing periodic boundary conditions, leading to the following system of coupled equations

\begin{equation}
\begin{split}
&\lambda_1=J_1 + \frac{1}{2\pi} \theta\left(\lambda_1 - \lambda_2\right) + \frac{1}{2\pi} \theta\left(\lambda_1 - \lambda_3\right), \\
&\lambda_2= J_2 + \frac{1}{2\pi} \theta\left(\lambda_2 - \lambda_1\right) + \frac{1}{2\pi} \theta\left(\lambda_2 - \lambda_3\right), \\
&\lambda_3= J_3 + \frac{1}{2\pi} \theta\left(\lambda_3 - \lambda_1\right) + \frac{1}{2\pi} \theta\left(\lambda_3 - \lambda_2\right),
\label{lambda_I}
\end{split}
\end{equation}
where $\theta(x) = -2 \arctan\left(x/c\right)$, and $J_1$, $J_2$ and $J_3$ are are distinct relative integers.

\section{Calculation of the matrix elements [Eq.(9)]}
We want to compute
\begin{equation}
M_{\vec\lambda,\vec\mu} = \langle \lambda_1, \ldots, \lambda_N | e^{-i K/\hbar_e \sum_j \cos x_j} | \mu_1, \ldots, \mu_N \rangle,
\end{equation}
namely
\begin{equation}
M_{\vec\lambda,\vec\mu} = \int \prod_j dx_j e^{-i K/\hbar_e\cos x_j} \Psi_{\lambda_1, \ldots, \lambda_N}^*(x_1, \ldots, x_N) \Psi_{\mu_1, \ldots, \mu_N}(x_1, \ldots, x_N),
\label{matrix-el2}
\end{equation}
with 
\begin{equation}
\Psi_{\lambda_1, \ldots, \lambda_N}(x_1, \ldots, x_N) = \sum_Q A_Q(\{\lambda_i\}, c) \sum_P \theta_P(x_1, \ldots, x_N) e^{i \sum_j x_{P(j)} \lambda_{Q(j)}},
\label{eq:wave}
\end{equation}
being the many-body wavefunction for $N$ Lieb-Liniger bosons  restricted to the fundamental rapidity sector $\lambda_1<\lambda_2<...<\lambda_N$ where $Q$ and $P$ are permutations of $S_N$.
Putting expression (\ref{eq:wave}) in Eq. (\ref{matrix-el2}), one obtains
\begin{equation}
M_{\vec\lambda,\vec\mu} = \sum_{Q,Q'} A_Q^* A_{Q'} \int \prod_j dx_j e^{-iK/\hbar_e \cos x_j} \sum_{P,P'} \theta_P(x_1, \ldots, x_N) \theta_{P'}(x_1, \ldots, x_N) e^{-i \sum_q (x_{P(q)} \lambda_{Q(q)} - x_{P'(q)} \mu_{Q'(q)})},
\end{equation}
with \(A_Q = A_Q(\{\lambda_i\}, c)\) and \(A_{Q'} = A_{Q'}(\{\mu_i\}, c)\) which are defined in Eq. (4) of the main text. This is only different from 0 when \(P = P'\), which leads to
\begin{equation}\label{eq_matel}
M_{\vec\lambda,\vec\mu} = \sum_{Q,Q'} A_Q^* A_{Q'} \sum_P \int_{\Gamma_P} \prod_j dx_j e^{-iK/\hbar_e \cos x_j} e^{-i \sum_q x_{P(q)} (\lambda_{Q(q)} - \mu_{Q'(q)})},
\end{equation}
with \(\Gamma_P\) delimiting the integration domain \(x_{P(1)} < \cdots < x_{P(N)}\). As every sector is equivalent, we can write the matrix element in a factorized form as
\begin{equation}
M_{\vec\lambda,\vec\mu} = \sum_{Q,Q'} \mathcal{A}_{Q,Q'} S_{Q,Q'},
\end{equation}
with $\mathcal{A}_{Q,Q'}=A_Q^* A_{Q'}$ and
\begin{equation}
S_{Q,Q'} = N! \int_{x_1 < \cdots < x_N} \prod_j dx_j e^{-iK/\hbar_e \cos x_j} e^{-ix_j (\lambda_{Q(j)} - \mu_{Q'(j)})}.
\end{equation}
While dealing with periodic boundary conditions (PBC), this formulation of the matrix element becomes ill-defined. The boundary condition imposes for every ordering to loop, which results in the integration domain having the form \(x_1 < \cdots < x_N < x_1\). Writing the sector this way may suggest that the integration domain has no dimension. However, we can overcome this issue by treating one of the variables (in this case, \(x_1\)) as a moving boundary for the other \((N - 1)\) variables. The corresponding domain is actually \(N\) times larger than \(x_1 < \cdots < x_N\) and leads to the ensuing expression
\begin{equation}
S_{Q,Q'} = N! \int_0^{L/\ell} dx_1 e^{-iK/\hbar_e \cos x_1} e^{-ix_1(\lambda_{Q(1)} - \mu_{Q'(1)})} \prod_{j=2}^{N} \int_{x_{j-1}}^{L/\ell} dx_j e^{-iK/\hbar_e \cos x_j} e^{-ix_j (\lambda_{Q(j)} - \mu_{Q'(j)})}.
\end{equation}
By using the following identity
\begin{equation}
e^{-iK/\hbar_e \cos x} = \sum_{n=-\infty}^{\infty} (-i)^n J_n(K/\hbar_e) e^{- i n x },
\end{equation}
the \(S\) function takes the form
\begin{equation}
S_{Q,Q'} = N! \sum_{\substack{m_1,\ldots,m_N }} (-i)^{\sum_j m_j} J_{m_1}(K/\hbar_e) \cdots J_{m_N}(K/\hbar_e) s_{Q,Q',\vec l},
\end{equation}
with
\begin{equation}
s_{Q,Q',\vec l} = \int_0^{2\pi} dx_1 e^{-il_1 x_1} \prod_{j=2}^{N} \int_{x_{j-1}}^{2\pi} dx_j e^{-il_j x_j},
\label{sqq1l}
\end{equation}
where
\begin{equation}
l_j = (\lambda_{Q(j)} - \mu_{Q'(j)}) + m_j.
\end{equation}

By performing the integration in Eq. (\ref{sqq1l}) for $N=3$, one obtains 
\begin{equation}
    \begin{split}
&s_{Q,Q',\vec l}
 = \frac{i e^{-i 2\pi (l_1 + l_2 + l_3)}}{l_1 l_2 (l_1 + l_2) l_3 (l_2 + l_3) (l_1 + l_2 + l_3)} \\
& \quad \left( e^{i 2\pi (l_1 + l_2 + l_3)} l_1 l_2 (l_1 + l_2) - l_2 l_3 (l_2 + l_3) + e^{i 2\pi l_1} (l_1 + l_2) l_3 (l_1 + l_2 + l_3) - e^{i 2\pi (l_1 + l_2)} l_1 (l_2 + l_3) (l_1 + l_2 + l_3) \right).\\
\end{split}
\label{sing}
\end{equation}

\begin{figure}
    \centering
    \includegraphics[width=0.5\linewidth]{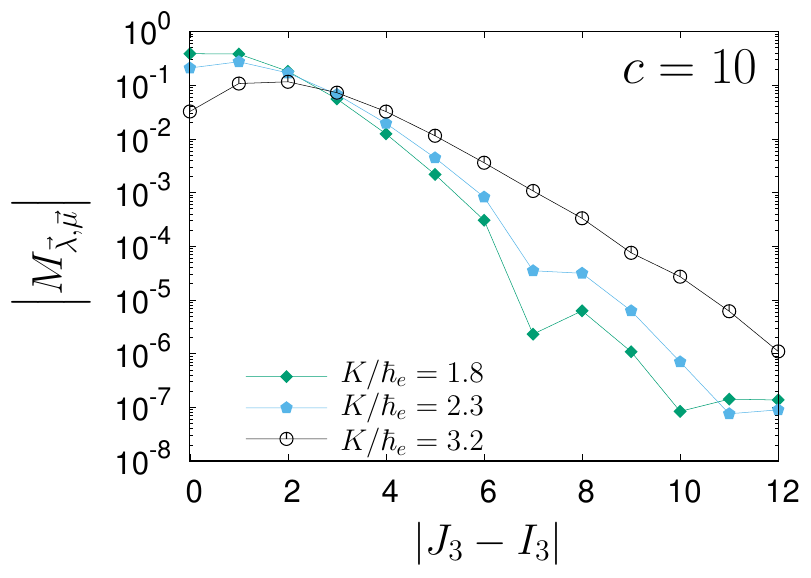}
    \caption{Matrix elements $|M_{\vec\lambda,\vec\mu}|$ for the case $c=10$, $I_1=J_1=-1$,
$I_2=J_2=0$, as functions of $|J_3-I_3|$, for different values of the stochasticity parameter $K/\hbar_e$.}
    \label{fig:mat_ele}
\end{figure}

Expression (\ref{sing}) has some artificial singularities that can be
eliminated by considering separately all the cases where the denominator vanishes as follows:

\begin{equation}
s_{Q,Q',\vec l} = 
\begin{cases} 
-2\pi e^{-i 2\pi (l_1 + l_2 + l_3)} \left( \frac{1}{l_2 + l_3} e^{i \pi l_1} \Phi_1 + \frac{1}{l_2 + l_3} e^{i \pi(l_1 + l_2)} \Phi_2 \right) & \text{if } l_3 \neq -l_2 \\ 
e^{-i 2\pi l_1}\left(\frac{ i e^{i 2\pi (l_1 + l_2)}}{l_2^2 (l_1 + l_2)} + \frac{ e^{i 2\pi l_1} (i l_2 + l_1 (-i + 2\pi l_2))}{l_1^2 l_2^2} - \frac{i}{l_1^2 (l_1 + l_2)}\right) & \text{if } l_3 = -l_2, \ l_1 \neq -l_2, \ l_1 \neq 0, \ l_2 \neq 0 \\ 
\frac{-i + i e^{i 2\pi l_2} + 2\pi l_2 (1+ i \pi l_2)}{ l_2^3} & \text{if } l_3 = -l_2, \ l_1 = 0, \ l_2 \neq 0 \\ 
\frac{i - i e^{-i 2\pi l_1} + 2\pi l_1 - i 2\pi^2 l_1^2}{ l_1^3} & \text{if } l_2 = 0, \ l_3 = 0, \ l_1 \neq 0 \\ 
\frac{-2\left(i + \pi l_1 + e^{-i 2\pi l_1} (-i + \pi l_1)\right)}{l_1^3} & \text{if } l_3 = -l_2, \ l_1 = -l_2, \ l_1 \neq 0 \\ 
\frac{4\pi^3}{3} & \text{if } l_1 = 0, \ l_2 = 0, \ l_3 = 0 
\end{cases}
\end{equation}

where

\begin{equation}
\Phi_1 = 
\begin{cases} 
\frac{\text{Sinc}(\pi l_1) - e^{ i \pi l_2} \text{Sinc}\left(\pi (l_1 + l_2)\right)}{l_2} & \text{if } l_2 \neq 0 \\ 
\frac{-\pi l_1 \cos(\pi l_1) + (1 - i \pi l_1) \sin(\pi l_1)}{\pi l_1^2} & \text{if } l_1 \neq 0 \\ 
-i \pi & \text{otherwise}
\end{cases}
\end{equation}

and

\begin{equation}
\Phi_2 = 
\begin{cases} 
\frac{-\text{Sinc}(\pi (l_1 + l_2)) + e^{ i \pi l_3} \text{Sinc}\left(\pi (l_1 + l_2 + l_3)\right)}{l_3} & \text{if } l_3 \neq 0 \\ 
\frac{\pi (l_1 + l_2) \cos(\pi (l_1 + l_2)) + (-1 +i \pi (l_1 + l_2)) \sin(\pi(l_1 + l_2))}{\pi (l_1 + l_2)^2} & \text{if } l_1 \neq -l_2 \\ 
i\pi & \text{otherwise,}
\end{cases}
\end{equation}
with $\textrm{Sinc}(z)=\sin(z)/z$. We show in Fig. \ref{fig:mat_ele} the behaviour of $|M_{\vec\lambda,\vec\mu}|$, where $\vec\lambda$ and $\vec\mu$ are determined
by the sets of relative integers $\vec I$ and $\vec J$ respectively (see Eq. (\ref{lambda_I})), for the case $c=10$, $I_1=J_1=-1$,
$I_2=J_2=0$, as functions of $|J_3-I_3|$. We can observe that for the range of values of the stochasticity parameter $K/\hbar_e$
that are relevant for this work, the matrix elements $|M_{\vec\lambda,\vec\mu}|$ decrease exponentially with respect the rapidities distance.

Remark that it would be possible to express the matrix elements $M_{\vec\lambda,\vec\mu}$ as functions of form factors \cite{Korepinetal,Slavnov,deNardisPanfil}. This would allow to have an analytical formula for the limiting case $1/c\ll 1$ and an arbitrary number of particles $N$ \cite{GranetEssler}, but the method proposed here is more efficient for the numerical calculation at any value of the interaction strength and for the specific case of $N=3$. We leave the larger $N$ case for future investigations.
\subsection{Condition for dimensionality reduction} Here we formally demonstrate that a system of $N$ kicked Lieb-Liniger bosons
reduces to $N$ kicked independent particles ($N$ independent 1D systems) only in the non-interacting regime ($c=0$) and in the fermionized regime
($c\rightarrow\infty$).
When this occurs, the matrix elements
\begin{equation}
    M_{\vec\lambda,\vec\mu} = \langle \lambda_1, \ldots, \lambda_N | e^{-i K/\hbar_e  \sum_j \cos x_j} | \mu_1, \ldots, \mu_N \rangle,
\end{equation}
can be rewritten as
\begin{equation}
    M_{\vec\lambda,\vec\mu} = \sum_{Q,Q'} \beta_{Q,Q'}\prod_{j=1}^N\langle \lambda_{Q(j)}| e^{-i K/\hbar_e \cos x_j} | \mu_{Q'(j)} \rangle,
\end{equation}
which explicitly factorizes in terms of single-particle contributions. The term $\beta_{Q,Q'}$ and the sum over $Q$ and $Q'$ account for the symmetry properties of the many-body wavefunction. Starting from expression \eqref{eq_matel} for the matrix elements, we have
\begin{equation}
M_{\vec\lambda,\vec\mu}=\sum_{Q,Q'} A_Q^* A_{Q'}\sum_P \int_{\Gamma_P}\!\!\prod_j dx_j e^{-iK/\hbar_e \cos x_j} e^{-i \sum_q x_{P(q)} (\lambda_{Q(q)} - \mu_{Q'(q)})}.
\end{equation}
Without loss of generality, we can set $j=P(q)$, leading to
\begin{equation}
M_{\vec\lambda,\vec\mu}=\sum_{Q,Q'} A_Q^* A_{Q'}\sum_P \int_{\Gamma_P}\!\!\prod_j dx_j e^{-iK/\hbar_e \cos x_j} e^{-i x_{j} (\lambda_{QP^{-1}(q)} - \mu_{Q'P^{-1}(q)})}.
  \label{pat1}
\end{equation}
Eq. (\ref{pat1}) simplifies to
\begin{equation}
M_{\vec\lambda,\vec\mu}=\sum_{\tilde Q,\tilde Q'} {A_{\tilde Q}}^* A_{\tilde Q'} \prod_j\int_{0}^{2\pi} dx_j e^{-i K/\hbar_e \cos x_j} e^{-i x_{j} (\lambda_{\tilde Q(q)} - \mu_{\tilde Q'(q)})}.
  \label{pat2}
\end{equation}
 where we define $\tilde Q=QP^{-1}$ and $\tilde Q'=Q'P^{-1}$. This factorization holds \emph{only if} 
$A^*_{\tilde Q}A_{\tilde Q'}=A^*_{Q}A_{ Q'}$ for all $Q, Q'$, and $P$, which translates into the condition
 \begin{equation}
 \begin{split}
   &  \prod_{1\le j,j'<k,k'\le N}\left(1\!-\!\dfrac{ic}{\lambda_{Q(j)}\!-\!\lambda_{Q(k)}}\right)\left(1\!+\!\dfrac{ic}{\mu_{Q'(j')}\!-\!\mu_{Q'(k')}}\right)\\
   &=\!\!\!\prod_{1\le j,j'<k,k'\le N}\left(1\!-\!\dfrac{ic}{\lambda_{\tilde Q(j)}\!-\!\lambda_{\tilde Q(k)}}\right)\left(1\!+\!\dfrac{ic}{\mu_{\tilde Q'(j')}\!-\!\mu_{\tilde Q'(k')}}\right).\\
     \end{split}
 \end{equation}
This equality holds for all $Q, Q'$, and $P$ only if $c=0$ or $c\rightarrow\infty$.

\section{Mapping to the Anderson model}
In this section we detail the procedure to get Eq. (7) of the main text from the Floquet equation
$U| \phi_\alpha\rangle=e^{-i \frac{\mu_\alpha}{\hbar_e}} |\phi_\alpha\rangle$, with $U=e^{- i H_0/\hbar_e}e^{- i H_K/\hbar_e}$.
By defining $|{\tilde  \phi}_{\alpha}\rangle = e^{iH_0/ 2\hbar_e} |\phi_{\alpha}\rangle$, one obtains
the time-reversal symmetrized Floquet equation 
\begin{equation}
\tilde U|\tilde \phi_\alpha\rangle=e^{-i \frac{\mu_\alpha}{\hbar_e}} |\tilde \phi_\alpha\rangle,
\label{sym1}
\end{equation} with 
\begin{equation}\tilde U=e^{- i H_0/(2\hbar_e)}e^{- i H_K/\hbar_e}e^{- i H_0/(2\hbar_e)}.
\end{equation} 
Furthermore, using the identity $\tilde U^\dagger U|\tilde \phi_\alpha\rangle=|\tilde \phi_\alpha\rangle$,
the following equation 
\begin{equation}
\tilde U^\dagger|\tilde \phi_\alpha\rangle=e^{i \frac{\mu_\alpha}{\hbar_e}} |\tilde \phi_\alpha\rangle
\label{sym2}
\end{equation}
holds too.
By projecting  Eqs. (\ref{sym1}) and (\ref{sym2}) onto the basis state $\langle\vec{\lambda}|$, one gets
\begin{equation}
  \label{eq_Ufloquet2}
\sum_{\vec{\nu},\vec{\mu},\vec{\beta}}  
\langle \vec{\lambda}|e^{\mp i H_0/(2\hbar_e)}|\vec{\nu}\rangle\langle\vec{\nu}| e^{\mp i H_K/\hbar_e}|\vec{\mu}\rangle\langle\vec{\mu}|e^{\mp i H_0/(2\hbar_e)}|\vec{\beta}\rangle\langle\vec{\beta}|{\tilde  \phi}_{\alpha}\rangle=e^{- i \frac{\mu_\alpha}{\hbar_e}} \langle \vec{\lambda}|{\tilde  \phi}_{\alpha}\rangle,
\end{equation}
namely
\begin{equation}
  \label{eq_Ufloquet3}
\sum_{\vec{\mu}}  
e^{\mp i (E_{\vec{\lambda}}+E_{\vec{\mu}})/(2\hbar_e)}\langle\vec{\lambda}| e^{\mp i H_K/\hbar_e}|\vec{\mu}\rangle\langle\vec{\mu}|{\tilde  \phi}_{\alpha}\rangle=e^{\mp i \frac{\mu_\alpha}{\hbar_e}} \langle \vec{\lambda}|{\tilde  \phi}_{\alpha}\rangle.
\end{equation}
By defining the matrix elements $M_{{\vec \lambda},{\vec \mu}}  = \langle{\vec \lambda} \vert e^{-i H_K/\hbar_e} \vert {\vec \mu} \rangle$
and by setting
\begin{equation}
 W_{{\vec \lambda},{\vec \mu}}  = 2 \, {\rm Re} \biggl\lbrace \exp\Bigl(-i \frac{ E_{\vec \lambda}+E_{\vec \mu}}{2\hbar_e}  \Bigr) M_{{\vec \lambda},{\vec \mu}}\biggr\rbrace,
 \end{equation}
 one gets straightforwardly Eq. (7) of the main text:
\begin{equation}
\sum_{{\vec \mu} \neq {\vec \lambda}} \! W_{{\vec \lambda},{\vec \mu}}  \langle {\vec \mu} \vert {\tilde \phi}_{\alpha}\rangle    +   W_{{\vec \lambda},{\vec \lambda}}  
\langle {\vec \lambda} \vert {\tilde \phi}_{\alpha}\rangle  =
 2 \cos\Bigl(\frac{\mu_{\alpha}}{ \hbar_e}\Bigr)  \langle {\vec \lambda} \vert {\tilde \phi}_{\alpha}\rangle.
\label{FloquetEq9}
\end{equation}
\section{Numerical procedure}

Once the matrix elements are obtained, the main computational effort is completed. The evolution operator over one period is given by:

\begin{equation}
    \hat{U} = e^{-i \frac{\hat{H}_0}{\hbar_e}} e^{-i \frac{\hat{H}_K}{\hbar_e}} .
\end{equation}
The matrix elements of the evolution operator in the Bethe basis $|\vec \lambda\rangle$ are:
\begin{equation}
        U_{\{\lambda_i\}, \{\mu_i\}} \equiv \langle \lambda_1, \ldots, \lambda_N | \hat{U} | \mu_1, \ldots, \mu_N \rangle,
\end{equation}
which can be written as:
\begin{equation}
    U_{\{\lambda_i\}, \{\mu_i\}} = e^{-i E_{\vec \lambda}/\hbar_e} \langle \lambda_1, \ldots, \lambda_N | e^{-i \frac{\hat{H}_K}{\hbar_e}} | \mu_1, \ldots, \mu_N \rangle.
\end{equation}
From our previous calculations, the matrix elements of the evolution operator are given by:
\begin{equation}
U_{\{\lambda_i\}, \{\mu_i\}} = e^{-i \frac{\hbar_e}{2} (\lambda_1^2 + \ldots + \lambda_N^2)} M_{\vec \lambda, \vec \mu}.
\end{equation}
We recall that \( M_{\vec \lambda, \vec \mu} \) is given by:
\begin{equation}
M_{\vec \lambda, \vec \mu} = \sum_{Q,Q'} A_{Q,Q'} N! \sum_{\substack{|m_1|,\ldots,|m_N| \leq M}} (-i)^{\sum_j m_j} J_{m_1}(\kappa) \cdots J_{m_N}(\kappa) s_{Q,Q'}(\{l_i\}).
\end{equation}
Here $M$ acts  as  a  cut-off  in  the  sum  over  Bessel  functions. According to asymptotic properties of Bessel functions we use $M>\lfloor 2K/\hbar_e \rfloor$ where $\lfloor \cdot \rfloor$ is the integer part. Since the full  Hilbert  space is infinite-dimensional,  we  truncate it to make numerical computations  feasible.  As explained  in  the  main  text,  we  restrict  Bethe  numbers  to  the  range $-N_s\le \lambda_i \le N_s$ with $N_s$ up to $29$. Furthermore, due to the symmetry of the wave function, it is sufficient to consider only ordered sets of Bethe numbers. This results in a matrix of size $N_m=\binom{2N_s+1}{3}$. Finally, we have to ensure that the matrix elements are properly normalized. Within the definition of the Bethe amplitude given in this appendix this is not the case. Proper normalization is restored by dividing each matrix elements by $\sqrt{\langle\vec\lambda|\vec\lambda\rangle\langle\vec\mu|\vec\mu\rangle}$. These norms can be directly evaluated using $M_{\vec \lambda, \vec \lambda}$ at $K=0$.

To compute the time evolution of the system, we expand its wave function in the Lieb-Liniger basis as:
\begin{equation}
|\psi(t)\rangle = \sum_{\{\lambda_i\}} \alpha_{\{\lambda_i\}}(t) |\lambda_1, \ldots, \lambda_N \rangle,
\end{equation}
where the coefficients \( \alpha_{\{\lambda_i\}}(t) \) evolve stroboscopically as:
\begin{equation}
\alpha_{\{\lambda_i\}}(t+1) = \sum_{\{\mu_i\}} U_{\{\lambda_i\}, \{\mu_i\}}(\kappa) \alpha_{\{\mu_i\}}(t),
\end{equation}
with $\alpha_{\{\lambda_i\}}(0)=1$ in the ground state and zero otherwise.

The total energy of the system at time \( t \) is given by:
\begin{equation}
\langle E(t)\rangle = \sum_{\{\lambda_i\}} | \alpha_{\{\lambda_i\}}(t) |^2 E_{\{\lambda_i\}},
\end{equation}
where \( E_{\{\lambda_i\}} = \frac{\hbar_e^2}{2} (\lambda_1^2 + \ldots + \lambda_N^2) \) is the energy associated with the quantum numbers.


\bibliographystyle{apsrev4-2}